\documentclass[runningheads]{llncs}
\usepackage[T1]{fontenc}

\usepackage{amsfonts}
\usepackage{amsmath}

\newcommand\numberthis{\addtocounter{equation}{1}\tag{\theequation}}
\usepackage{graphicx}

\begin{document}

\title{Addressing the Subsumption Thesis: A Formal Bridge between Microeconomics and Active Inference}
\titlerunning{Addressing the Subsumption Thesis}

\author{Noé Kuhn\ }%\and 

\institute{University of St. Gallen}

\maketitle              % typeset the header of the contribution
\begin{abstract}
As a unified theory of sentient behaviour, active inference is formally intertwined with multiple normative theories of optimal behaviour. Specifically, we address what we call the subsumption thesis: The claim that expected utility from economics, as an account of agency, is subsumed by active inference. To investigate this claim, we present multiple examples that challenge the subsumption thesis. To formally compare these two accounts of agency, we analyze the objective functions for MDPs and POMDPs. By imposing information-theoretic rationality bounds (ITBR) on the expected utility agent, we find that the resultant agency is equivalent to that of active inference in MDPs, but slightly different in POMDPs. Rather than being strictly resolved, the subsumption thesis motivates the construction of a formal bridge between active inference and expected utility. This highlights the necessary formal assumptions and frameworks to make these disparate accounts of agency commensurable.

\end{abstract}

\section{Introduction}

Since the middle of the previous century, expected utility has formed the bedrock of the agency underwriting microeconomics. With early implementations dating back to Bernoulli in 1713 \cite{ref_szipro}, expected utility has undergone many augmentations in order to reflect realistic deliberate decision processes. The comprehensive start of this lineage can be traced to the classic utility theorem \cite{ref_vNM}; \cite{ref_bonanno}, with earlier applications found in \cite{ramsey1926truth}. Subsequent accounts include Bayesian Decision Theory \cite{ref_savage1972foundations}; \cite{berger1985statistical}, Bounded Rationality \cite{ref_simon1957models}; \cite{ref_ortega2015}, Prospect Theory \cite{ref_kahneman1979prospect}, and many more flavours. The algorithmic implementation of expected utility theory is found in the Reinforcement Learning literature \cite{ref_bartosutton}. While seemingly disparate, practically all expected utility accounts of agency depict an agent making decisions in a probabilistic setting to attain optimal reward -- to pursue utility \cite{ref_bentham1781introduction}. \\
\indent Coming from the completely different background of neuroscience, Active Inference a comparatively new account of agency \cite{ref_friston2009free}, positioning itself as ``a unifying perspective on action and perception [...] richer than the common optimization objectives used in other formal frameworks (e.g., economic theory and reinforcement learning)'' \cite[pg. 1;4]{ref_pezzulo2024sentient}. Here the agent seeks to minimize information-theoretic surprisal expressed as free energy (See Definition 3, 4). Active inference allows for a realistic modeling of the very neuronal processes underwriting biological agency \cite{ref_parr2019neuronal}.\\
\indent Given the breadth of successful applications \cite{ref_dacosta2020discretesynthesis} combined with its strong fundamental first principles \cite{ref_friston2023simple}, some proponents of active inference posited what we call the Subsumption thesis: Expected utility theory as seen in economics is subsumed by active inference -- it is an edge case. A formulation in the same vein posits: ``Active inference [...] englobes the principles of expected utility theory [...] it is theoretically possible to rewrite any RL algorithm [...] as an active inference algorithm'' \cite{ref_dacosta2024agency}. So how does the subsumption thesis hold up in the given examples? Is it possible to formally delineate how expected utility and active inference differ? This paper then establishes a firm connection between microeconomics and active inference, which has scarcely been explored before \cite{ref_henriksen2020variational}.\\
\indent To formally compare the two accounts of agency, we require a commensurable space for agent-environment interactions: MDPs and POMDPs (Definition 1 and 2). These agent-environment frameworks are the bread and butter of expected utility applications \cite{ref_bellman1957markovian}, \cite{ref_bartosutton}, \cite{ref_littman}. Active inference agency has more recently also been specified for the same frameworks  \cite{ref_fristonbook}, \cite{ref_dacosta2020discretesynthesis}, \cite{ref_dacosta2022reward}, \cite{ref_dacosta2024agency}. As such (PO)MDPs provide a theoretical arena for the subsumption thesis to be evaluated.  \\
\indent What exactly is at stake that motivates this inquiry into the subsumption thesis? Firstly, expected utility and active inference rest upon different first principles to substantiate their respective account of agency \cite{ref_dacosta2024agency}. Analysis of the formal relationship between these two accounts could provide insights into how the first principles of one account might be a specification the other's first principles. Secondly, this inquiry will shed light on how each account handles the exploration-exploitation dilemma \cite{ref_berger-tal2014exploration}: How should an agent prioritise between exploring an environment versus exploiting what they already know about the environment for utility? Finally, if active inference truly subsumed expected utility, then the ramifications for welfarist economics would be enormous: Currently, the formal mainstream understanding of welfare which informs economic policy \cite{ref_pigou1920economics} is based on aggregating individual agents acting according to expected utility \cite[pg. 45]{ref_mongin2006concept} \cite{ref_ross2014philosophy}. The subsumption thesis challenges foundations of `optimal' economic policy if expected utility only captures a sliver of `optimal' behaviour.\\
\indent To investigate the subsumption thesis, the rest of the paper is structured as follows. In section \textbf{2.} the agent-environment frameworks are defined alongside the relevant accounts of agency and basic concepts in microeconomics. In section \textbf{3.}, some examples are investigated which challenge the subsumption thesis. In section \textbf{4.}, the formal bridge between expected utility and active inference is established via Information Theoretic Bounded Rationality (ITBR) \cite{ref_ortega2015}. Finally section \textbf{5.} provides some concluding and summarizing remarks.\\

\section{Preliminary Definitions and Microeconomics}

\subsection{Agent-Environment Frameworks}

A finite Markov Decision Process (MDP) is a mathematical model that specifies the elements involved in agent-environment interaction and development \cite{ref_bartosutton}. This formalization of sequential decision making towards reward maximization originates in dynamic programming, and currently enjoys much popularity in model-based Reinforcement Learning (RL). Although potentially reductive, employing MDPs and POMDPs allows for formal commensurability between different accounts of agency.\\ 

\noindent \textbf{Definition 1} (Finite Horizon MDP). \textit{An MDP is defined according to the following given tuple: $(\mathbb{S},\mathbb{A},P(s'|a,s),R(s',a),\gamma = 1, \mathbb{T})$}
\begin{itemize}
    \item $\mathbb{S}$ is a finite set of states.
    \item $\mathbb{A}$ is a finite set of actions.
    \item $P(s'|a,s)$ is the transition probability of posterior state $s'$ occurring upon the agent's selection of action $a$ in the prior state $s$.
    \item $R(s',a) \in \mathbb{R}^+$ is the reward function taking as arguments the agent's action and resulting state. For our purposes, the action taken will be irrelevant to the resulting reward: $R(s',a) = R(s')$.
    \item $\gamma$ denotes the discount factor of future rewards. This is set to $1$ as this parameter is not commonly used in the cited active inference literature.
    \item $\mathbb{T} = \{1,2,\ldots,t, \ldots,\tau, \ldots, T\}$ is a finite set for discrete time periods whereby $t < \tau$ and the horizon is $T$.
\end{itemize}
\noindent Note that time period subscripts e.g, $s_\tau$ are sometimes omitted when unnecessary.\\

\noindent In a single-step decision problem, an expected reward-maximizing agent would evaluate the optimal action $a^*$ as follows:

\begin{equation} \label{reward max agent}
a_{t}^{*} = \underset{a \in \mathbb{A}}{\arg\max} \ E_{P(s_\tau|a_t,s_t)} R(s_\tau)
\end{equation}

\noindent Further, a Partially Observable Markov Decision Process (POMDP) generalizes an MDP by introducing observations $o$ that contain incomplete information about the latent state $s$ of the environment \cite{ref_littman,ref_bartosutton}. The agent can only infer latent states via observations. Thus, POMDPs are ideal for modeling action-perception cycles \cite{ref_friston2023simple} with the cyclical causal graphical model $a \rightarrow s \rightarrow o \ldots  $  \\

\noindent \textbf{Definition 2} (Finite Horizon POMDP). \textit{A finite horizon POMDP further adds two elements to the previously given MDP tuple: $( \mathbb{O} , P(o|s) )$ }
\begin{itemize}
    \item $\mathbb{O}$ is a finite set of observations.
    \item $P(o|s)$ is the probability of observation $o$ occuring to the agent given the state $s$.
\end{itemize}

\subsection{Active Inference Agency}

\noindent With the environment-agent frameworks established, we can proceed to define how an active inference agent approaches a (PO)MDP. Although fundamental and interesting, the Variational Free Energy objective crucial to perception in active inference will not be examined here; Inference on latent states is assumed to occur through exact Bayesian inference \cite[pg. 16]{ref_dacosta2022reward}. The central objective function for agency in active inference is the Expected Free Energy (EFE), the formulation of which for (PO)MDPs we will take from \cite{ref_dacosta2020discretesynthesis},\cite{ref_dacosta2024agency},\cite{ref_dacosta2022reward},\cite{ref_fristonbook}. Essentially, the agent takes the action trajectory  $\pi = \{a_\tau, \ldots, a_T\}$ that minimizes the cumulative expected free energy $G$, which is roughly the sum of the single-step EFEs $G_\tau$. By inferring the resultant EFE of policies through $Q(\cdot)$, the optimal trajectory $\pi^*$ corresponds to the most likely trajectory -- the path of least action. \cite{ref_friston2023simple}. Formally: 
\begin{align*} \numberthis
    & \pi^* = \underset{\pi}{\arg\min} \  G(\pi)    \\
   &  G(\pi) \approx \sum\limits_{\tau}^{T} G_\tau(\pi) \\
    & G_\tau(\pi) = G_(a_t)
\end{align*}
We can then define the EFE for single-step for MDPs and POMDPs. Note that this could also be scaled up to trajectories/vectors of the relevant elements e.g $s_{t:T}$. For simplicity we will look at single-step formulations for the remainder of the paper.  \\

\noindent \textbf{Definition 3:} (EFE on MDPs). \textit{For an agent in an MDP with preference distribution $P(s|C)$, the Expected Free Energy of an action for some given current state $s_t$ is defined as follows:}
\begin{align*} \numberthis \label{EFE MDP}
   & G_\tau(a_t) =  D_{KL}[P(s_\tau|a_t, s_t)||P(s|C)] \\
    & = -\underbrace{\mathfrak{H}[P(s_\tau|a_t, s_t)]}_\text{Entropy of future states} \underbrace{- E_{P(s_\tau|a_t)}[logP(s|C)]}_\text{Expected Surprise}
\end{align*}
\noindent As seen in the rearranged objective function of the second line, the agent seeks to keep future options open while meeting preferences; The entropy of future possible states is to be maximized while the information-theoretic surprisal according to the preference distribution is to be minimized. The conditionalisation on $C$ specifies a parameterized preference distribution \cite{ref_fristonbook}. \\

\noindent \textbf{Definition 4:} (EFE in POMDPs). \textit{For an agent in a POMDP with preference distribution $P(s|C), P(o|C)$, the Expected Free Energy of an action for some given current state $s_t$ is defined as follows:}\\
\begin{align*}\numberthis \label{EFE POMDP}
& G_\tau(a_t) = \underbrace{E_{P(s_\tau|o_t, a_t)} \mathfrak{H}[P(o_\tau|s_\tau)]}_\text{Ambiguity} + \underbrace{D_{KL}[P(s_\tau|a_t, o_t)|P(s|C)]}_\text{Risk}   \\ 
& = - \underbrace{E_{P(o_\tau|a_t)}[D_{KL}[P(s_\tau|o_\tau)||P(s_\tau|a_t)] ]}_\text{Intrinsic Value} - \underbrace{E_{P(o_\tau, s_\tau|a_t)}[logP(o|C)]}_\text{Extrinsic Value}
\end{align*}
\noindent With some auxiliary assumptions \cite[pg. 10]{ref_dacosta2024agency} which are admissible for our purposes, the two formulations of EFE in a POMDP are equivalent, and both contain a curiosity inducing term and an exploitation term \cite{ref_friston2015epistemic}. The first formulation motivates the agent to minimize the expected entropy of observations given unknown states and to minimize the divergence between actual states and preferred states. The second formulation motivates the agent to maximize the expected informational value of observations while also maximizing the expected log probability of preferred observations -- note how the underbrace does not include the minus.
\subsection{Microeconomics}
\indent As this paper investigates an intersection between fields which are generally not in direct contact, a brief introduction to risk attitudes and lotteries in microeconomics is provided. The origin of these studies can be traced back to the gambling houses of the 18th century; As early as 1713, Bernoulli employed marginally decreasing utility functions to resolve the famous St. Petersburg Paradox \cite{ref_szipro}. This paradox asks what amount a rational agent would be willing to pay to enter lottery with an infinite expected value. To answer this question, we utilize lotteries \cite{ref_bonanno} and risk-attitudes \cite{ref_arrow} from microeconomics:\\
\noindent \textbf{Definition 5:} (Lottery). \textit{A (monetary) lottery is a probability distribution over outcomes $x$ that are the argument of the utility function. Therefore, a lottery $L$ can be modelled as an integrable random variable defined by the  probability space triplet consisting of a sample space, sigma algebra, and probability measure: $(\Omega, \mathfrak{F}, \mu)$ }

\noindent A decision maker then evaluates their preference over a set of lotteries according to their utility function $U(x) \in \mathbb{R}^+$ where $x \in \mathfrak{F}$. The expected utility of each lottery then induces a preference ordering over lotteries. For example, the strong preference relation $L_1 \succ L_2$ means that lottery $L_1$ is more preferable to lottery $L_2$.  Classically, this ordering is in line with the von Neumann-Morgenstern axioms of completeness, transitivity, continuity, and independence \cite{ref_vNM}. Any such preference ordering is also maintained for any positive affine transformation of $U(x)$ \cite{ref_bonanno}; \cite{ref_vNM}. By juxtaposing the expected utility $E[U(L)]$ of a lottery against the utility of the expectation of the same lottery $U(E[L])$, risk aversion can be defined.\\
\noindent \textbf{Definition 6:} (Risk Aversion). \textit{An agent with some utility function $U(\cdot)$ is considered risk averse if for some lottery the following preference relation holds: $U(E[L]) \succ U(L)$ }\\

\noindent This preference relation occurs if an agent's utility function is concave, i.e the marginal utility is decreasing. A risk loving agent conversely acts according to a convex utility function, and a risk neutrality is associated with a linear utility function. Accordingly, Bernoulli used lotteries and a log-utility function to resolve the St. Petersburg paradox, the solution of which is relegated to Appendix A for readers unfamiliar with the problem -- the pertinent point is that concave utility functions on set rewards are extensively studied in economics.

\section{Subsumption Examples}

Equipped with an understanding of marginal utility and lotteries, we can now tackle two manifest exhibits of the subsumption thesis by proponents of active inference. Further, an illustrative MDP demonstrates the divergence in behaviour between active inference and expected utility. The results of the simulated behaviour are directly taken from the discussed papers. These exhibits then motivate the bridging in section \ref{bridge} later. \\
\indent The first \cite{ref_sajid2022active} and second \cite{ref_dacosta2024agency} exhibit both concern agency in a classical T-maze: A simple forked pathway in which the agent can either go left or right (See Figure 1 below). This environment is also called ``Light'' in POMDP literature \cite{ref_littman}. The agent-environment dynamics are modeled using a POMDP; Unbeknownst to the agent, the reward is either in the left or right arm. The agent can also go down to observe a cue indicating the definite location of the reward. Going down the `wrong' arm of the fork leads to a punishment equal to the negative reward, say $-1$. The performance of the agency is evaluated by the reward attainment of the agent within a two period horizon. At this point however, the setup of the first and second exhibit diverge crucially.

\begin{figure}
    \centering
    \includegraphics[width = 3cm, height = 3cm]{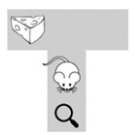}
    \caption{An agent in a T-Maze with unknown context. Illustration from \cite{ref_sajid2022active}}
    \label{Agency in a T-Maze}
\end{figure}

In the first exhibit \cite{ref_sajid2022active}, the right and left fork are absorbing states -- the agent cannot leave them upon entry. As such, the agent cannot correct going down the wrong arm in the first period by the second period. Given this setup, the expected utility agent performs very poorly, while the active inference is cue-seeking and therefore performs optimally \cite[pg. 138]{ref_sajid2022active}. The expected utility agent performs so poorly because supposedly ``the agent does not care about
the information inferred and is indifferent about going to the cue location
or remaining at the central location''\cite[pg. 137]{ref_sajid2022active}. This appears reductive, as an expected utility agent facing two lotteries will behave the same as the active inference agent. Consider the risky lottery $L_1$ which is the result of a gambling and non-information seeking strategy. Contrast this lottery with $L_2$, which is the degenerate lottery of investigating the cue first and going to the reward in the second period. Assuming even just a linear utility function $U(R) = R(s)$, then $U(L_1) = 0.5\cdot 1 + 0.5\cdot -1 = 0$ and $U(L_2) = 1 \cdot 1$. Clearly, the expected utility agent holds a preference which motivates cue-seeking behaviour: $L_2 \succ L_1 $.\\ 
\indent Regarding the second exhibit \cite{ref_dacosta2024agency}, there is a slight difference in the setup. The arms of the fork are no longer absorbing states, which allows for mistake correction and a cumulative reward of $2$ over two periods. Now, the focus of \cite{ref_dacosta2024agency} isn't anymore on performance comparison but instead achieving the desiderata of risk-aversion and information sensitivity \cite[pg. 10]{ref_dacosta2024agency}. While the agency according to active inference meets the desiderata, the expected utility agent does not. However, risk aversion and the resulting information sensitivity can easily be induced by using a concave utility function. Consider again the risky lottery $L_1$ and a cue-seeking lottery $L_2$. Assuming a utility function taking the reward as argument $U(R) = R(s)^c$ where $c \in \mathbb{R}^+$, then $U(L_1) = 0.5 \cdot 0 + 0.5 \cdot 2^c $ and $U(L_2) = 1^c$. Accordingly if $c < 1$, then $L_2 \succ L_1$, and only if $c = 1$, then indeed the agent is indifferent $L_1 \sim L_2$. As is evident, it is the risk-neutral agent who does not meet the desiderata. \\
\indent Finally, consider the following single-step MDP created for illustrative purposes. A paraglider stands at the foot of two steep mountains $s_1,s_2$ separated by a chasm $s_3$ and must decide which one to climb. While still risky, the path up mountain $1$ is far more secure than the path up mountain $3$. However, mountain $2$ is taller than mountain $1$ and therefore allows for a more enjoyable flight. This decision process can aptly be modeled in an MDP (See Figure 2 below). Note that the subscript here does not relate to the period. Taking $a_1$ gives $\{P(s_1|a_1), P(s_2|a_1), P(s_3|a_1) \} = \{0.6, 0, 0.4\}$, and $a_2$ gives $\{P(s_1|a_2), P(s_2|a_2),$ $ P(s_3|a_2) \} = \{0, 0.4, 0.6\}$. The height in kilometers gives the reward function $\{R(s_1), R(s_2), R(s_3)\}$ $ = \{1, 1.5, 0\}$. With the MDP sufficiently specified, we can compare the agency of an active inference agent and an expected utility agent. See Appendix B for details on the resulting expected utility and free energy.\\
\begin{figure}
    \centering
    \includegraphics[width = 7.5cm, height = 5cm]{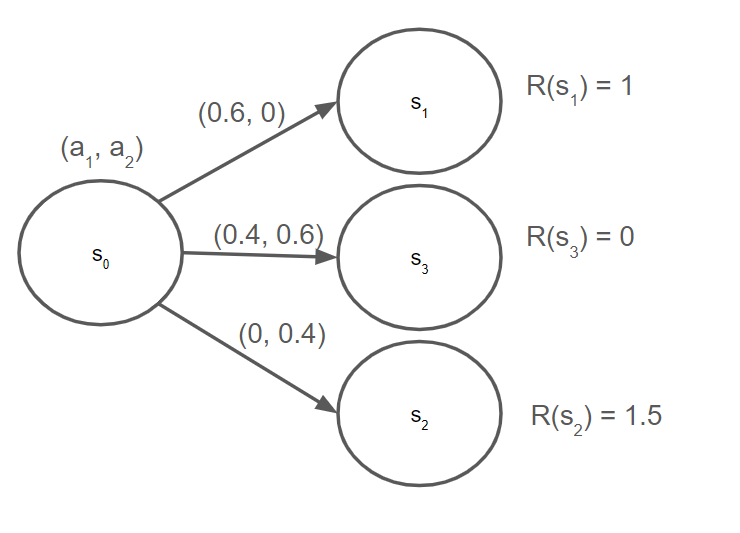}
    \caption{`Paraglider' MPD with states, actions, transition probabilities, and rewards}
    \label{fig:enter-label}
\end{figure}
\indent The active inference agent is indifferent between the two actions as both actions result in the same EFE -- equation (\ref{EFE MDP}). For expected utility however, only the linear utility function agent is indifferent; The risk averse agent prefers the safer mountain and the risk loving agent prefers the riskier mountain due to the concavity or convexity of the utility function respectively. As such, this simple but valid MDP provides a setup in which specified expected utility may better meet the desiderata than the active inference agent.

It should be clear by now that wrapping a utility function around the rewards is a well-studied and principled approach which differs from simply including ``ad-hoc exploration bonuses in the reward function'' \cite[pg. 2]{ref_dacosta2024agency}. Introducing non-linearity over the rewards seems to lead to an impasse in the comparison between expected utility. The most direct case for comparing and subsuming expected utility \cite{ref_dacosta2022reward} only considers a linear utility function ($U(\cdot) = R(\cdot)$) for the expected utility agent. Even if non-linearity for expected utility were considered in \cite{ref_dacosta2022reward}, it appears unclear to us as to how the resulting agency -- more specifically the induced careful and explorative aspect -- could be compared in a generalized manner.  \\
To resolve this issue of incommensurability, we would like to draw attention to physical and biological constraints on agents which have motivated active inference. For example, tractability is a central concern for active inference as evidenced by the appeal to variational Bayes. Luckily, there already exists an account of agency which imbues expected utility with constraints: ITBR \cite{ref_ortega2015},\cite{ref_braun2015hierarchy}. The connection between ITBR and active inference in an MDP has briefly been explored before \cite{ref_ortega2015epistemic}. We seek to now clearly establish this conceptual bridge between microeconomics and active inference for both MDPs and POMDPs.

\section{From expected utility to active inference via ITBR} \label{bridge}

\subsection{In MDPs}

Let us first establish the bridge between expected utility and active inference in an MDP. Essentially, both objective functions can both be transformed into the ``Divergence Objective'' \cite{ref_millidge2021info}: 
\begin{equation} \label{divergence obj. mdp}
a^* = \underset{a \in \mathbb{A}}{\arg\min} D_{KL}[P(s_\tau|a_t)||P^*(s)]
\end{equation}
Where $a^*$ is the optimal action and $P^*(s)$ is a preference distribution over states, for example, a softmax or Gibbs distribution. Note the immediate similarity to the EFE objective function for MDPs (\ref{EFE MDP}) -- here conditionalisation on the current state $s_t$ is omitted for brevity as we consider a single step. \\  
\indent To get there from expected utility, we can consider the following
Lagrangian constraints on the utility objective function \cite[pg. 3]{ref_braun2015hierarchy}. Let $P(\cdot)$ be the prior distribution over relevant elements of the MDP, and $Q(\cdot)$ the posterior distribution after a limited search or `bounded deliberation; see \cite{ref_ortega2015} for details. The deliberation bound is given as an information-theoretic quantity e.g $nats$ or $bits$; hence the name information-theoretic bounded rationality. Let $K \in \mathbb{R}^+ \ nat$ -- although the information theoretic unit base $nat$ is arbitrary:
\begin{align*} 
    & D_{KL}[Q(s_\tau|a_t)||P(s_\tau|a_t)] \leq K  \numberthis \label{constraint ITBR} 
\end{align*}
The constraint of equation \ref{constraint ITBR} can be interpreted as a bound on the search for the optimal action. The second constraint of (\ref{constraint ITBR}) means that the agent is uncertain about the `true' transition probabilities in the MDP. This constraint gives us the following ITBR free energy objective function \cite{ref_ortega2015epistemic}: 
\begin{equation} \label{F ITBR objective}
    F_{ITBR}(Q) = \sum\limits_s Q(s|a) \ \left( U(s,a) - \frac{1}{\beta} log \frac{Q(s|a)}{P(s|a)} \right)
\end{equation}
This functional is to be maximized ($Q^*(s|a)$) with given parameter $\beta \in \mathbb{R}^+$. See Appendix C for how the maximizing solution is derived.  
We can now use the maximizing argument of the objective function (\ref{F ITBR objective}) as a `goal' for the agent, or a preference distribution over states $P^*(s)$. Like in active inference, we assume that the preference distribution over states is independent of the action taken to get there. This preference is given by the Gibbs distribution: 
\begin{equation} \label{gibbs preference MDP}
    P^*(s|a) = \frac{P(s|a) \cdot e^{\beta U(s,a)}}{Z_\beta} \rightarrow P^*(s)
\end{equation}

\noindent We can now solve (\ref{gibbs preference MDP}) for $U(s,a)$ and input this into (\ref{F ITBR objective}) to obtain the divergence objective (\ref{divergence obj. mdp}): 
\begin{equation} \label{final ITBR MDP objective}
    a^* = \underset{a \in \mathbb{A}}{\arg\min} - D_{KL}[Q(s_\tau|a_t)||P^*(s)] + constant
\end{equation}
Where the constant is irrelevant for optimization purposes. The details of this derivation relegated to Appendix D. Evidently, the same optimal agency arises in an MDP for an active inference and ITBR agent. Next, let us bridge expected utility to active inference in a POMDP.

\subsection{In POMDPs}

Analogously to the MDP setting, we can transform the ITBR objective to get to the divergence objective function for POMDPs. Fortunately, this divergence objective has previously been formulated as the ``Free Energy of the Expected Future'' (FEEF): \cite[pg. 10]{ref_millidge2021whence}. Again, this objective function motivates a minimal posterior divergence from a preference distribution, now jointly over states and observations:
\begin{equation} \label{divergence obj. POMDP}
   a^* = \underset{a \in \mathbb{A}}{\arg\min} D_{KL}[P(o_\tau,s_\tau|a_t)|| P^*(o,s)]
\end{equation}

\noindent To attain this expression, we can formulate a new ITBR objective in the POMDP framework \cite{ref_braun2015hierarchy} and transform it analogously to the MDP case before. We can again consider the information-theoretic bound $V\in \mathbb{R}^+ nat$  :
\begin{align*} 
   D_{KL}[Q(s_\tau, o_\tau|a_t)||P(s_\tau, o_\tau|a_t)] \leq V  \numberthis 
\end{align*}
\noindent Considering these constraints, we can express the ITBR Free energy objective function again:
\begin{align*} 
    F_{ITBR} (Q) = \sum \limits_s Q(o,s|a) \left( U(o,s,a) - \frac{1}{\beta} log \frac{Q(o,s|a)}{P(o,s|a)}   \right)  \numberthis \label{F ITBR POMDP}
\end{align*}
\noindent Where the solution is again the Gibbs distribution: 
\begin{align*}
    P^*(o,s|a) = \frac{P(o,s|a) e^{\beta U(o,s,a)}}{Z_\beta}
    \numberthis \label{Gibbs preference POMDP}
\end{align*}

\noindent By combining (\ref{Gibbs preference POMDP}) and (\ref{F ITBR POMDP}) we get the resultant minimization objective, where the resultant optimal agency is of course the same as that of the divergence minimization objective (\ref{divergence obj. POMDP}):
\begin{equation}
   a^* = \underset{a \in \mathbb{A}}{\arg\min} - D_{KL} [Q(o_\tau,s_\tau|a_t)||P^*(o,s)] + constant
\end{equation}
Which again intuitively motivates the agent to have the inferred posterior distribution given the action be as close as possible to the prior preference distribution over states. It is crucial to note however that this is not the same objective function as EFE in POMDPs (\ref{EFE POMDP})! To get from the divergence objective (\ref{divergence obj. POMDP}) for POMDPs to EFE (\ref{EFE POMDP}), we can follow the steps taken in \cite{ref_millidge2021whence}; for a detailed discussion of the relationship between the divergence objective and EFE, the reader should also consult \cite{ref_millidge2021info}, \cite{ref_millidge2021whence}. Essentially, the divergence objective can also be decomposed into an exploitative and explorative term. However, while the explorative term is equal to that of active inference, the divergence objective additionally further encourages the agent to increase posterior entropy of observations given latent states -- to keep options open. Note that in the formulation below, both objective functions below (\ref{IBTR POMDP decomposed}), (\ref{EFE POMDP}) are to be minimized. 
\begin{align*} \numberthis \label{IBTR POMDP decomposed}
& - F_{ITBR} = D_{KL}[Q(o,s|a)||P^*(o,s)]\\
& = \underbrace{E_{Q(s|a)} \left[ \  D_{KL}[Q(o|s)||P^*(o)] \  \right]}_\text{Extrinsic Value} - \underbrace{E_{Q(o|a)}\left[ \  D_{KL}[Q(s|o)||Q(s|a)] \ \right] }_\text{Intrinsic Value} 
\end{align*}
\begin{align*}\tag{\ref{EFE POMDP}}
& G \ = \ -\underbrace{E_{Q(o, s|a)}[logP(o|C)]}_\text{Extrinsic Value} - \underbrace{E_{Q(o|a)}\left[ \ D_{KL}[Q(s|o)||Q(s|a)] \ \right]}_\text{Intrinsic Value}  
\end{align*}

Whereby the relationship between $G$ and $-F_{ITBR}$ is as follows:

\begin{equation} \label{diff. G and F}
    G - E_{Q(o|a)} \mathfrak{H}[Q(o|s)] = -F_{ITBR}
\end{equation}

\noindent Comparing then the decomposed divergence objective to active inference, in the pursuit of extrinsic value the boundedly rational utility agent seeks to additionally keep posterior options open compared to the active inference agent -- similarly to the agency in an MDP. 

\subsection{The Bridge summarized}

Let us reconsider the entire journey from expected utility to active inference so as to not lose sight of the forest in front of all the trees. First, simply incorporate a utility function into the reward maximizing objective function (\ref{reward max agent}) to get an expected utility agent. Then, impose information-theoretic deliberation constraints on the optimization process (\ref{constraint ITBR}). Consequently, the agent faces a Lagrangian optimization problem (\ref{F ITBR objective}). The solution to this optimization problem is taken as a preference distribution for the agent. Combining the preference distribution and the objective function results in the divergence objective \cite{ref_millidge2021info}, which can then be compared with the active inference objective function. In an MDP, the resultant agency is the exact same (\ref{divergence obj. mdp}). However, in a POMDP, the objective functions differ (\ref{diff. G and F}). \\
Bar this difference, one key aspect must be elucidated for the extrinsic value terms in both MDPs and POMDPs. Although the intrinsic value term is the same for the different objective functions, the two prior preference distributions $P^*(s)$ of ITBR (\ref{gibbs preference MDP}) and $P(s|C)$ of active inference (\ref{EFE MDP}) are not necessarily the same. For $\beta = 1$, if we consider $P(s|C)$ as a Gibbs distribution as per \cite[pg. 9]{ref_dacosta2022reward};\cite[pg. 134]{ref_fristonbook}, then the two preference distributions are only equal if either the utility function is linear, or if active inference admits agent-specific utility functions (17); An admission which prima facie seems irreconcilable with the physicalist/nonsubjectivist philosophy behind active inference. This larger discussion is, however, to be relegated to a later paper.
\begin{equation}
P^*(s) = \frac{e^{U(s)}}{ \sum \limits_s e^{U(s)} }
   \quad\mathrm{and}\quad 
P(s|C) = \frac{e^{R(s)}}{ \sum \limits_s e^{R(s)} }
\end{equation}
Where optimal behaviour in an MDP, i.e $a^*$, is the same for both accounts of agency only if $U(s)$ is a positive affine transformation of $R(s)$.

\section{Conclusion}

Having formalized the bridge from expected utility to active inference, we can re-evaluate the subsumption thesis. Simple reward-oriented agency ($U(\cdot) = R(\cdot)$) can be effectively subsumed by active inference in MDPs, and if exact Bayesian inference is used, also in POMDPs \cite{ref_dacosta2022reward}. However, as shown in section \textbf{2.}, expected utility in microeconomics uses utility functions that take rewards as arguments. As seen in section \textbf{3.} then, there are various examples where the subsumption argument does not hold up; Expected utility acts the same as active inference, or under specific circumstances, may meet desiderata of agency even more. In section \textbf{4.}, we establish the formal bridge between expected utility and active inference. By using ITBR \cite{ref_ortega2015}, we can directly compare the objective functions of (bounded) expected utility and active inference. Upon considering agent-environment assumptions, the divergence objective \cite{ref_millidge2021info} is used as a reference point to compare the two accounts of agency. It is demonstrated that in an MDP, ITBR and active inference lead to the same agency \cite{ref_ortega2015epistemic}. In a POMDP, ITBR is equivalent to the divergence objective, which however differs from the active inference objective function \cite{ref_millidge2021whence}. While the explorative/information-seeking terms are equal, the exploitative/reward-oriented term differs: $E_{Q(o|a)} \mathfrak{H}[Q(o|s)]$ must be subtracted from the active inference objective function, and the preference distributions are not necessarily equal.\\
\indent An area where expected utility cannot compete however is in the first principles which motivate agency \cite{ref_friston2009free}, \cite{ref_friston2023simple} \cite{friston2022path}, \cite{ref_barp2022geometric}, \cite{ref_dacosta2024agency}. Still, the debate on what objective function follows from the first principles is not yet sealed in this flourishing field \cite{ref_millidge2021whence}. Perhaps more intriguing links between brain function and the physical interpretations of information theory lurk underneath the bridge established here. Furthermore, computational simulations \cite{ref_braun2015hierarchy} and empirical studies \cite{schwartenbeck2015evidence} might flesh out the practical comparison between bounded expected utility and active inference; Computational efficiency has not remotely been addressed in this paper. Finally, it would be especially interesting for economics to understand how an economy could develop from multiple ITBR or active inference agents \cite{hylandforthcomingmulti}. By integrating interdisciplinary approaches to agency, we aim to foster a holistic understanding of agency that enriches the roles of both human and artificial agents in society.

\newpage

\begin{credits}
\subsubsection{\ackname}  I would like to express my immense gratitude to my supervisor for allowing me to delve into this topic and lending his support along the way. Further, I want to thank the various researchers willing to so openly discuss the contents and concepts of the paper. Only thanks to those fruitful exchanges could these connections across varying fields even be grasped.

\subsubsection{\discintname}
The author has no competing interests to declare that are
relevant to the content of this article.
\end{credits}\\

%
% ---- Bibliography ----

%\bibliographystyle{splncs04}
%\bibliography{Paper}

%
%\begin{thebibliography}{8}

%\bibitem{ref_article1}
%Author, F.: Article title. Journal \textbf{2}(5), 99--110 (2016)

%\bibitem{ref_lncs1}
%Author, F., Author, S.: Title of a proceedings paper. In: Editor,
%F., Editor, S. (eds.) CONFERENCE 2016, LNCS, vol. 9999, pp. 1--13.
%Springer, Heidelberg (2016). \doi{10.10007/1234567890}

%\bibitem{ref_book1}
%Author, F., Author, S., Author, T.: Book title. 2nd edn. Publisher,
%Location (1999)

%\bibitem{ref_proc1}
%Author, A.-B.: Contribution title. In: 9th International Proceedings
%on Proceedings, pp. 1--2. Publisher, Location (2010)

%\bibitem{ref_url1}
%LNCS Homepage, \url{http://www.springer.com/lncs}, last accessed 2023/10/25
%\end{thebibliography}
\newpage

\appendix
\renewcommand{\thesection}{Appendix}
\section{} % Leaving this empty will just display "Appendix A"
\section*{A:} % For parts within the appendix if needed, displaying as "A"
\textbf{Resolving the St. Petersburg Paradox}\\

\noindent Consider a lottery on the outcome of a fair coin toss. Starting at two dollars, the stake doubles with every subsequent outcome of heads. The game ends once tails comes up for the first time in the sequence. The expected payout $E[L]$ of the game is thus infinite:
$$
E[L] = \sum \limits_{i = 1}^\infty \frac{1}{2^i}\cdot 2^i = \infty
$$
How much would someone pay to participate in this game?  Taking a linear utility function on the payout, the gambler should be willing to pay any amount to enter the game. Daniel Bernoulli suggested a logarithmic utility function $U(x) = ln(x)$. Assume the cost of entry is $x$. Then the expected utility of the lottery has a finite value; The amount the agent at most would be willing to enter the lottery: 
$$
E[U(L)] = \sum \limits_{i = 1}^\infty \frac{1}{2^i}\cdot ln(2^i) = 2\cdot ln(2)
$$
Therefore the agent expects finite utility from the payout of the lottery due to the concavity of the utility function. As such, only a finite amount will be paid to enter the game.

\section*{B:}

\textbf{Expected Utility and Active Inference for the 'Paraglider' MDP}

\noindent The single-step MDP is specified as follows. Therefore note that the subscript does not pertain to the period: 
\begin{align*}
& \mathbb{S} = {s_1,s_2,s_3} \\
& \mathbb{A} = {a_1, a_2} \\
&\{P(s_1|a_1), P(s_2|a_1), P(s_3|a_1) \} = \{0.6, 0, 0.4\}\\
&\{P(s_1|a_2), P(s_2|a_2), P(s_3|a_2) \} = \{0, 0.4, 0.6\} \\
&\{R(s_1), R(s_2), R(s_3)\} = \{1, 1.5, 0\}
\end{align*}

\noindent Consider an expected utility agent with utility function $U(R(s)) = R(s)^c$ where $c \in \mathbb{R}^+$. As such, 
\begin{align*}
    & E[U(a_1)] = 0.6\cdot 1^c \\
    & E[U(a_2)] = 0.4 \cdot 1.5^c \\
    & \text{For $c < 1$} \rightarrow \underset{a \in \mathbb{A}}{\arg\max} \ E[U(a)] = a_1 \\
    & \text{For $c >1$} \rightarrow \underset{a \in \mathbb{A}}{\arg\max} \ E[U(a)] = a_2
\end{align*}
So a risk-averse expected utility agent will scale the smaller but safer mountain.\\

\noindent The active inference agent however is indifferent between the two actions. If we assume the preference distribution to be a softmax on the rewards, then we can ignore the normalizing denominator as it is constant w.r.t to action. Therefore we can write the relevant objective function as: 
\begin{align*}
& G(a_t) = -\sum\limits_s P(s_\tau|a_t)\cdot R(s_\tau) - \sum\limits_s P(s_\tau|a_t) log \frac{1}{P(s_\tau|a_t)} \\
& G(a_1) = - 0.6 - 0.3065 - 0.366 = G(a_2)\\
& \rightarrow \ \underset{a \in \mathbb{A}}{\arg\min} G(a) = \{a_1,a_2\}
\end{align*}
Therefore the optimal action of the risk-averse expected utility agent is a subset of the optimal active inference agency.\\

\section*{C: }

\textbf{Preference distribution derivation}\\

\noindent We maximize the ITBR objective function (\ref{F ITBR objective}) via first order condition.

\begin{align*}
& \frac{\delta F_{ITBR}}{ \delta Q(s|a)} = \ U(s,a) - \frac{1}{\beta} \left( log \frac{Q(s|a)}{P(s|a)} + 1\right) \stackrel{!}{=} 0
\end{align*}
Solve for $Q(s|a)$, and normalize to attain the Gibbs distribution
\begin{align*}
& Q(s|a) = P(s|a)e^{\beta U(s,a)-1} \propto P(s|a)e^{\beta U(s,a)} \\
& Q^*(s|a) = \frac{P(s|a)e^{\beta U(s,a)}}{\sum \limits_s P(s|a)e^{\beta U(s,a)} } = \frac{P(s|a)e^{\beta U(s,a)}}{Z_\beta}
\end{align*}

\noindent Which gives us (\ref{gibbs preference MDP})

\section*{D: }

\textbf{Getting from ITBR to the divergence objective via the Gibbs distribution.}\\

\noindent Solve (\ref{gibbs preference MDP}) for $U(s,a)$:

\begin{align*}
& P^* (s|a) = \frac{P(s|a) e^{\beta U(s)}}{Z_\beta} \\
&\frac{1}{\beta} ln(P^*(s|a) \cdot Z_\beta ) = U(s)
\end{align*}

\noindent Plug this into the ITBR objective function (\ref{F ITBR POMDP}) and consider the maximizing argument $a$: 

\begin{align*}
& \underset{a \in \mathbb{A}}{\arg\max} \  \frac{1}{\beta} E_{Q(s|a)} [lnP^*(s|a) + ln(Z_\beta)] - \frac{1}{\beta} ln \frac{Q(s|a)}{P(s|a)} \\
&= \underset{a \in \mathbb{A}}{\arg\max} \ E_{Q(s|a)} [lnP^*(s|a) + ln(Z_\beta) - ln Q(s|a) + ln P(s|a)] \\
&=  \underset{a \in \mathbb{A}}{\arg\min} \ E_{Q(s|a)} [-lnP^*(s|a) - ln(Z_\beta) + ln Q(s|a) - ln P(s|a)] \\
&= \underset{a \in \mathbb{A}}{\arg\min} \ D_{KL} [Q(s|a) || P^*(s|a)]
\end{align*}

\noindent Which is the divergence objective for MDPs (\ref{divergence obj. mdp}). In a POMDP setting, the derivation proceeds analogously to obtain the Free Energy of the Expected Future (\ref{divergence obj. POMDP}).

\end{document}